\documentclass[fleqn,usenatbib]{mnras}

\usepackage[T1]{fontenc}

\DeclareRobustCommand{\VAN}[3]{#2}
\let\VANthebibliography\thebibliography
\def\thebibliography{\DeclareRobustCommand{\VAN}[3]{##3}\VANthebibliography}

\usepackage{graphicx}	
\usepackage{amsmath}

\usepackage{amssymb}	
\usepackage{soul}
\usepackage{newtxtext,newtxmath}

\title[Multi-frequency studies of B0919+06 and B1859+07]{Multi-frequency study of the peculiar pulsars PSR~B0919+06 and PSR~B1859+07}

\author[Rajwade et al.]{K.~M.~Rajwade,$^{1}$\thanks{E-mail:kaustubh.rajwade@manchester.ac.uk}
B.~B.~P.~Perera,$^{2}$\thanks{E-mail:bhakthiperera@gmail.com}
B.~W.~Stappers,$^{1}$
J.~Roy,$^{3}$
A.~Karastergiou$^{4,5}$ and
\newauthor J.~M.~Rankin$^{6}$
\\
$^{1}$Jodrell Bank Centre for Astrophysics, University of Manchester, Oxford Road, Manchester M13 9PL, UK\\
$^{2}$Arecibo Observatory, University of Central Florida, HC3 Box 53995, Arecibo, PR 00612, USA\\
$^{3}$National Centre for Radio Astrophysics, University of Pune Campus, GaneshKhind, PO Box. , Pune, India\\
$^{4}$Astrophyiscs, Denys Wilkinson Building, University of Oxford, Keble Road, Oxford OX1 3RH, UK\\
$^{5}$Department of Physics and Electronics, Rhodes University, PO Box 94, Grahamstown 6140, South Africa\\
$^{6}$Physics Department, University of Vermont, Burlington, VT 05405, USA\\
}

\date{Accepted XXX. Received YYY; in original form ZZZ}

\pubyear{2015}

\begin{document}
\label{firstpage}
\pagerange{\pageref{firstpage}--\pageref{lastpage}}
\maketitle

\begin{abstract}
Since their discovery more than 50 years ago, broadband radio studies of pulsars have generated a wealth of information about the underlying physics of radio emission. 
In order to gain some further insights into this elusive emission mechanism, we performed a multi-frequency study of two very well-known pulsars, PSR~B0919+06 and PSR~B1859+07. These pulsars show peculiar radio emission properties whereby the
emission shifts to an earlier rotation phase before returning to the nominal emission phase in a few tens of pulsar rotations
(also known as `swooshes'). We confirm the previous claim that the emission during the swoosh is not necessarily absent at
low frequencies and the single pulses during a swoosh show varied behaviour at 220~MHz. We also confirm that in PSR~B0919+06,
the pulses during the swoosh show a chromatic dependence of the maximum offset from the normal emission phase with the offset
following a consistent relationship with observing frequency. We also observe that the flux density spectrum of the radio
profile during the swoosh is inverted compared to the normal emission. For PSR~B1859+07, we have discovered a new mode of
emission in the pulsar that is potentially quasi-periodic with a different periodicity than is seen in its swooshes. We invoke an emission model previously proposed in the literature and show that this simple model can explain the macroscopic observed characteristics in both pulsars. We also argue that pulsars that exhibit similar variability on short timescales may have the same underlying emission mechanism.  
\end{abstract}

\begin{keywords}
stars: neutron -- pulsars: general -- pulsars: individual 
\end{keywords}



\section{Introduction}
Pulsars are rotating neutron stars that emit radio waves from their magnetic poles. Since their discovery five decades ago~\citep{hewish1968}, more than 3000 have been discovered in our Galaxy, in globular clusters, and the neighbouring Magellanic Clouds\footnote{\url{https://www.atnf.csiro.au/research/pulsar/psrcat/}}~\citep{manchester2005,titus2020, Ye2019}. Though pulsars have a very stable pulse profile over a timescale of hours, peculiar variations in pulsed emission have also been observed on the timescales of a single rotation period. In short, three primary peculiarities in radio pulses from neutron stars are: 1) Nulling; 2) Mode-changing; and 3) Intermittency. Nulling is the cessation of radio emission for a few to many pulse periods~\citep{wang2007} while mode-changing pertains to a sudden change in the average pulse profile of the pulsar that can typically last for a few minutes to hours before switching back to the original pulse shape~\citep{rankin1986}. Over the years, many authors  have suggested that nulling and mode-changing are related phenomena where nulling can be thought of as an extreme mode change and are manifestations of a change in the magnetosphere of the neutron star~\citep{wang2007}.~\cite{lyne2010} showed a direct connection between radio emission variability and the spin-down rate in a sample of mode-changing pulsars suggesting a global change in the magnetosphere that alters the emission mechanism resulting in radio emission and spin-down torque changes. An extreme case of the nulling/mode-changing pulsars are intermittent pulsars~\citep[e.g.][]{kramer2006}. These pulsars switch from a normal emission state to a null state where no radio emission is seen from them. The time scale of these mode changes can last anywhere from a few weeks to a few years. Only five intermittent pulsars have been found to date~\citep{lyne2017}. Similar to other mode-changing pulsars, it was shown that the spin-down rate is smaller in the `null' state compared to the normal state~\citep[see][for more details]{kramer2006} in intermittent pulsars suggesting that both these phenomena are related and can be attributed to a global change in the magnetosphere of the neutron star.

Then there are pulsars like PSR B0611+22 that show mode-changing on extremely short timescales (a few minutes). The mode-changing is also very peculiar where the radio emission in the abnormal mode is offset in phase compared to the normal mode with negligible changes to the overall pulse shape unlike what is seen in canonical mode-changing pulsars~\citep{rajwade2016}. Similarly, there are pulsars that show episodic events in their emission whereby emission from subsequent rotations appears earlier in pulse phase for a few tens of pulse periods before returning back to the expected rotation phase. These have been given different names in the literature, but we will use `swooshes'~\citep{rankin2006} in this paper. PSRs B0919+06 and B1859+07 show this behaviour and have been studied in great detail over the last decade~\citep{rankin2006, wahl2016, shaifullah2017, han2016}. ~\cite{wahl2016} performed a detailed analysis of the swooshes at multiple frequencies to show that the swooshes seen in PSR B1859+07 are quasi-periodic in nature.~\cite{perera2015} and~\cite{perera2016} performed a detailed analysis of the long term spin-down behaviour of PSRs B0919+06 and B1859+07 to show that they exhibit two states in their spin-down rates, and there is no correlation between the swooshes and the spin-down state transitions suggesting that the swooshing is a manifestation of small scale variations in the current sheet in the magnetosphere. Recently, interest in these two pulsars has renewed with new observations over a large radio frequency range. Using observations over multiple frequencies, ~\cite{yu2019} and~\cite{shaifullah2017} showed that the swooshes of PSR B0919$+$06 were chromatic in nature whereby during a swoosh, the offset from the typical phase becomes smaller and emission itself becomes weaker at lower radio frequencies. All these observations pose important questions about how this behaviour can fit in the standard pulsar radio emission framework~\citep{Ruderman1975}, how these swooshing pulsars are related to other nulling and mode changing pulsars and how these different phenomena can be attributed to the same underlying emission mechanism. 

Here, we present an attempt to decipher the underlying physics of this peculiar radio emission with a simultaneous, multi-frequency radio campaign of PSRs B0919+06 and B1859+07. The paper is organised as follows: in section~\ref{obs} we describe the observations. In Section~\ref{results}, we show the results from the analysis that we performed on the dataset. We discuss the implications of our results in the context of pulsar emission models in Section~\ref{dis} and conclude with our findings in Section~\ref{sum}.

\section{Observations and data processing}
\label{obs}

We observed PSR B0919+06 simultaneously at 125$-$250 (band-2), 300$-$500 (band-3) and 550$-$750 (band-4) MHz using the upgraded Giant Metrewave Radio Telescope (uGMRT; \citep{gupta2017}). The usable bandwidth for band-2 was reduced to 60 MHz (i.e. 190$-$250 MHz), and for band-3 it was reduced to 140 MHz (removing the MUOS satellite affected channels and some other RFI contaminated channels). The wide instantaneous frequency coverage was achieved by splitting antennas into three sub-arrays using the multi-subarray observing mode supported in the GMRT Wide-band Backend (GWB; \citep{Reddy2017}). The antenna based amplitude and phase offsets were calibrated out in order to form three coherent array beams one for each frequency band. The number of antennas used to form beams were 5, 8 and 14 in band-2, band-3 and band-4 respectively. The GWB total-intensity 8-bit filterbank outputs having 4096$\times$ 0.0488 MHz from all the three beams were recorded at 327.68 $\mu$s time resolution. The scans on pulsars were interleaved with calibrator observations every 2 h for re-phasing the array, which were required to optimise the coherent array sensitivity. For PSR B0919+06, the phasing calibrator was 0837$-$198. In order to avoid degradation of sensitivity the relatively high DM pulsar B1859+07 closer to the Galactic plane was only observed at 550--750~MHz with the uGMRT. The calibrator used for amplitude and phase calibration in the observations of PSR B1859+07 was 1822$-$096. For the single sub-array band-4 observations, a total of 20 antennas were used to form a coherent array beam at the same time and frequency resolution used for PSR B0919+06.

Both PSRs B0919+06 and B1859+07 are regularly observed at $\sim$1.4~GHz as part of the pulsar timing program at the 76-m Lovell radio telescope in the UK. The Lovell observations of these pulsars were scheduled so as to obtain maximum overlap with the uGMRT observations. The two pulsars were observed on three epochs that matched the epoch of observations for the uGMRT. The data were recorded simultaneously with the Digital FiterBank (DFB) backend~\citep{manchester2013} at 1520~MHz over a bandwidth of 384~MHz and at a higher time resolution using a ROACH-based~\footnote{\url{https://casper.berkeley.edu}} backend at 1532~MHz~\citep{bassa2016}. For each observation, a polyphase filter coarsely channelised a 400~MHz band into 25 subbands of 16~MHz each. Each 16~MHz subband was further channelised into $32 \times 0.5$~MHz channels using \texttt{digifil} from the \texttt{dspsr} software suite~\citep{vanstraten2011}, and downsampled to a sampling time of 256~$\upmu$s. The 800 total channels, spanning 400~MHz, were then combined  in frequency. These were then reduced to 672 frequency channels over a bandwidth of 336~MHz in order to mitigate the effect of Radio Frequency Interference (RFI). We also masked frequency channels in the data containing persistent narrow-band RFI. The resulting fiterbank data were then saved to disk. The DFB data of PSR B0919$+$06 were saved after integrating 10 seconds of data folded to the best known period, while for the ROACH backend, we saved data for every pulse. However, for PSR 1859$+$07, we used the ROACH data that were coherently dedispersed and folded at the dispersion measure (DM) and period from the best known pulsar ephemeris. Only the resulting folded data integrated over 10-seconds were saved in order to get enough Signal-to Noise ratio (S/N) in every sub-integration to study the swooshes. The data were cleaned to remove strong RFI, and the times-of-arrivals generated from both backends were cross-checked as a sanity check for data quality and time keeping. 

In addition to the GMRT and Lovell telescope observations, PSR B1859$+$07 was observed with the 305-m Arecibo telescope in Puerto Rico on November 1, 2019 (MJD 58788) for 15 minutes. We used the L-wide receiver at 1380~MHz and a usable bandwidth of 600~MHz, and recorded the coherently dedispersed data with the PUPPI backend at a sampling rate of 10.24~$\mu$s. For the analysis presented below, we also used a few archival Arecibo telescope observations of the pulsar. These observations were obtained on MJDs 52739, 53372, 56377, and 57121 with durations of 10, 20, 70, and 130 min, respectively. The earlier two observations were made with the WAPP instrument that recorded four 100~MHz channels centred at 1275, 1425, 1525, and 1625~MHz. The later two observations were obtained using the Mock spectrometer and recorded four 86~MHz wide bands centred at 1270, 1420, 1520, and 1620~MHz. Both WAPP and Mock spectrometer data were recorded at a sampling rate of 512~$\mu$s. Since the swooshing time-scale of these pulsars is short ($\sim$ a few pulse periods), they can only be studied properly with single pulse data. Thus, we used the Lovell telescope's ROACH data of PSR B0919$+$06, the Arecibo telescope's data of PSR B1859$+$07, and the GMRT data of both pulsars in our analysis. The details of the observations are presented in Table~\ref{psr_info}

For each dataset, we dedispersed the data at the best known pulsar DM, and the resulting dynamic spectrum for each single pulse was saved to a file using the standard processing tools in \textsc{psrchive}~\citep{hotan2004}. Then, the data that were corrupted by RFI were excised using \textsc{clfd}~\citep{Morello2019} that performs time and frequency domain RFI excision of folded pulsar data. Finally, the cleaned datasets were collated together to create 10 minute long datasets for further analysis. 

\begin{table*}
\begin{center}
\caption{
Details of the observations used in this analysis. 
The last column indicates the number of swooshes that was identified using a filtering method based on the criteria of defining a swoosh as described in section 3.
GWB stands for GMRT Wideband backend~\citep{Reddy2017}, ROACH stands for the data acquisition pipeline based on ROACH field programmable gated array (FPGA) boards~\citep{bassa2016}. 
 }
\label{psr_info}
\begin{tabular}{lccccccc}
\hline
\multicolumn{1}{c}{PSR} &
\multicolumn{1}{c}{Telescope} &
\multicolumn{1}{c}{Backend} &
\multicolumn{1}{c}{Center Frequency} &
\multicolumn{1}{c}{Bandwidth} &
\multicolumn{1}{c}{MJD} &
\multicolumn{1}{c}{Obs. length} &
\multicolumn{1}{c}{No. of swooshes} \\
\multicolumn{1}{c}{ } &
\multicolumn{1}{c}{} &
\multicolumn{1}{c}{} &
\multicolumn{1}{c}{(MHz)} &
\multicolumn{1}{c}{(MHz)} &
\multicolumn{1}{c}{} &
\multicolumn{1}{c}{(min)} &
\multicolumn{1}{c}{} \\
\hline
\hline
B0919$+$06 & GMRT & GWB & 220/400/650 & 60/140/200 & 58455 & 150 & 3 \\
 & GMRT & GWB & 220/400/650 & 60/140/200 & 58461 & 150 & 8 \\
 & GMRT & GWB & 220/400/650 & 60/140/200 & 58467 & 150 & 6 \\
 & Lovell & ROACH & 1532 & 384 & 58455 & 167 & 2  \\
 & Lovell & ROACH & 1532 & 384 & 58461 & 178 & 5 \\
 & Lovell & ROACH & 1532 & 384 & 58467& 118 & 4\\
\hline
B1859$+$07 & GMRT & GWB & 650 & 200 & 58455 & 140 & 56 \\
 & GMRT & GWB & 650 & 200 & 58468 & 160 & 60 \\
 & GMRT & GWB & 650 & 200 & 58494 & 165 & -- \\
 & Arecibo & PUPPI & 1380 & 600 & 58788 & 15 & 10 \\
 & Arecibo & Mocks & 1270/1420/1520/1620 & 86 & 57121 & 130 & 129 \\
 & Arecibo & Mocks & 1270/1420/1520/1620 & 86 & 56377 & 70 & 53 \\
 & Arecibo & WAPP & 1275/1425/1525/1625 & 100 & 53372 & 20 & 13 \\
 & Arecibo & WAPP & 1275/1425/1525/1625 & 100 & 52739 & 10 & 8 \\
\hline
\end{tabular}
\begin{tabular}{l}
Note: The Mock and WAPP instruments recorded the data with a given bandwidth centered at four different frequencies.
\end{tabular}
\end{center}
\end{table*}

\section{Results}
\label{results}

For both pulsars, we separated out the swooshes in our dataset using the following methodology. In order to identify a swoosh in our dataset,  we first created a template profile to represent the normal emission mode in these pulsars. Since we were looking for any deviations in the pulse phase compared to the expected pulse phase of the normal emission profile, we created a folded pulse profile from a time-contiguous segment of data where no offset in the rotation phase was visually seen. Then, to generate a template, we fitted the resulting profile with a set of von Mises functions using \textsc{paas} which is part of the \textsc{PSRCHIVE} software suite~\citep{hotan2004}. Then, we cross-correlated each single pulse with our template to get an offset in phase for each pulse with respect to the phase corresponding to the peak in the template profile~\citep[see][for more details on the methodology]{Oslowski2011}. For any canonical pulsar that shows random pulse phase jitter in single pulses, the resulting phase offset distribution should be a Gaussian centred on zero but swooshes cause the distribution to become skewed or bimodal (depending on the number of swooshes) with a peak centred around the maximum phase offset during a swoosh. We defined the start of a swoosh as any pulse where the offset in phase compared to the template is more than twice the standard deviation of the phase offset distribution between the template and the actual data and the large deviation prevails for more than five consecutive pulses. The value of five was chosen based on the fact that in a Gaussian distribution of offsets, it is extremely unlikely to get 5 successive values deviant by more than 2-$\sigma$ from the mean of the distribution. We considered a swoosh to end once we encountered the first pulse which had shifted back in phase so that it no longer satisfied the criterion of being offset by twice the standard deviation of the phase offset distribution between the template and the actual data. A swoosh then refers to the collection of pulses within the start and end points. Using the criteria described above, we obtained a total of 17 swooshes from 3 observing epochs of B0919+06 and more than a 100 swooshes for PSR~B1859+07 (see Table~\ref{psr_info}).

\begin{figure*}
\centering
	\includegraphics[scale=0.42]{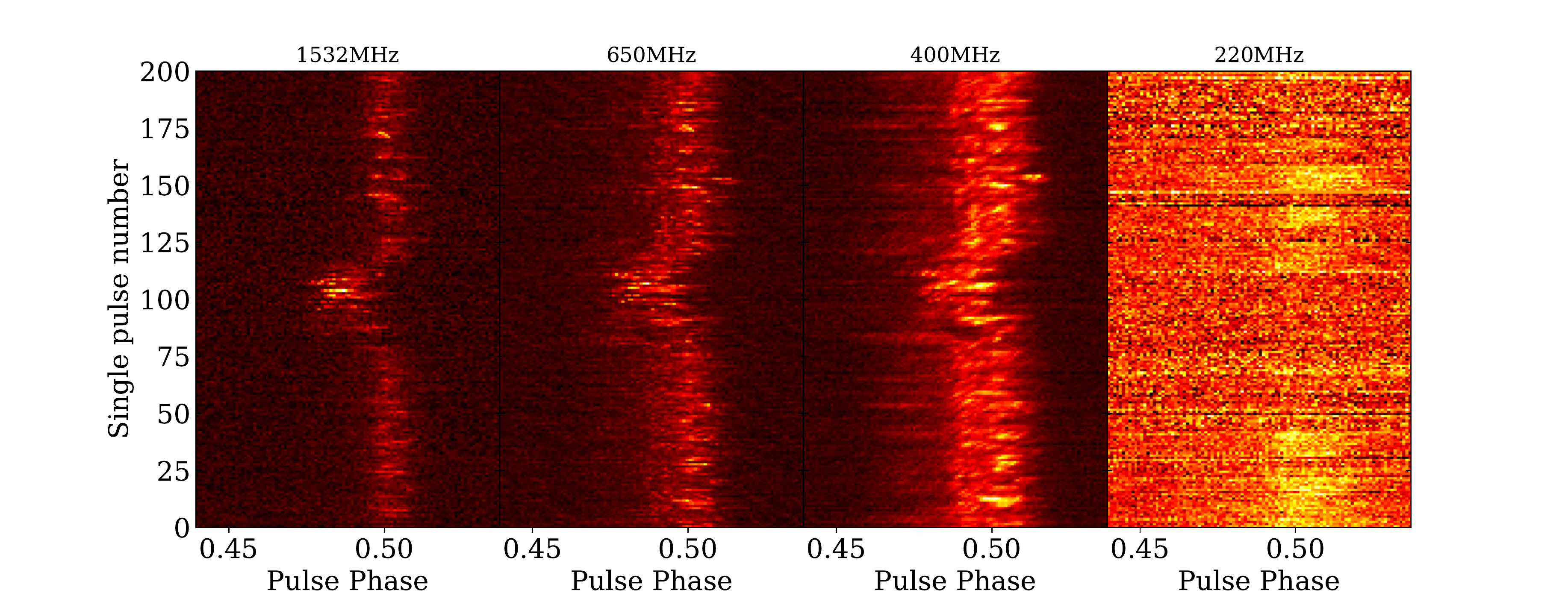}
		\includegraphics[scale=0.42]{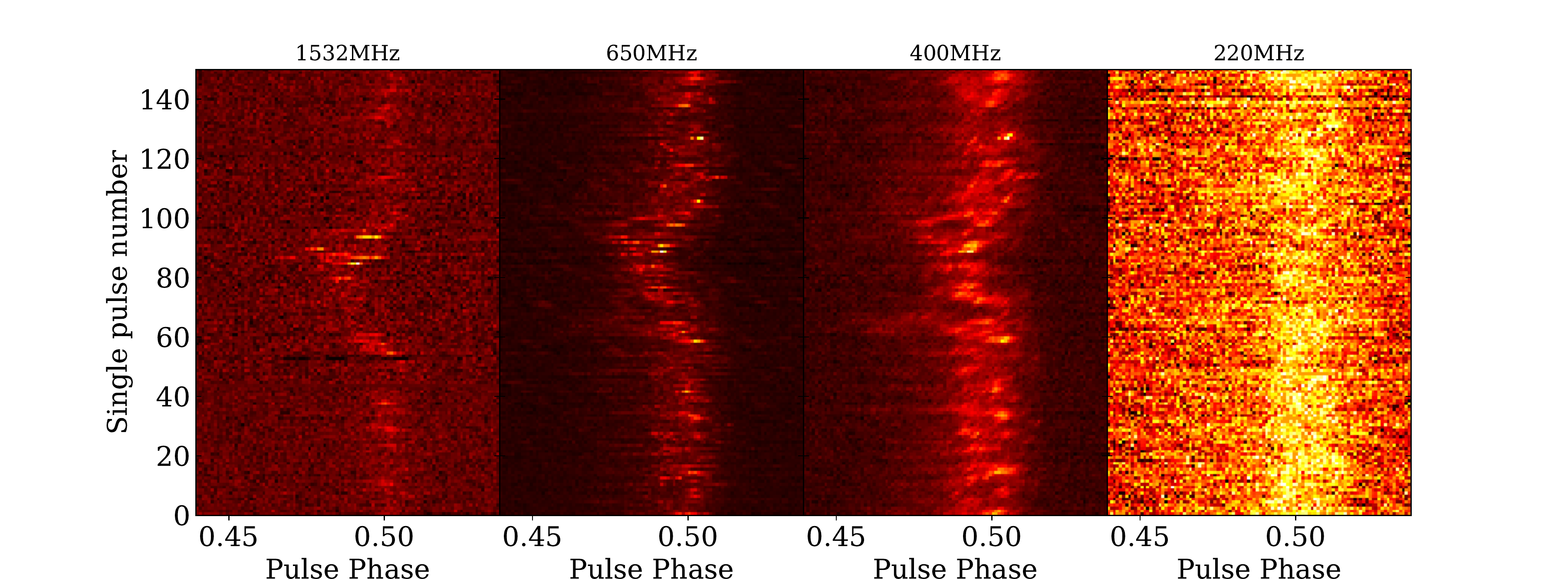}
    \caption{Example swooshing events from PSR B0919+06 observed simultaneously at four radio frequencies revealing strong frequency dependence.}
    \label{fig:b0919_example}
\end{figure*}

\begin{figure}
\includegraphics[scale=0.65]{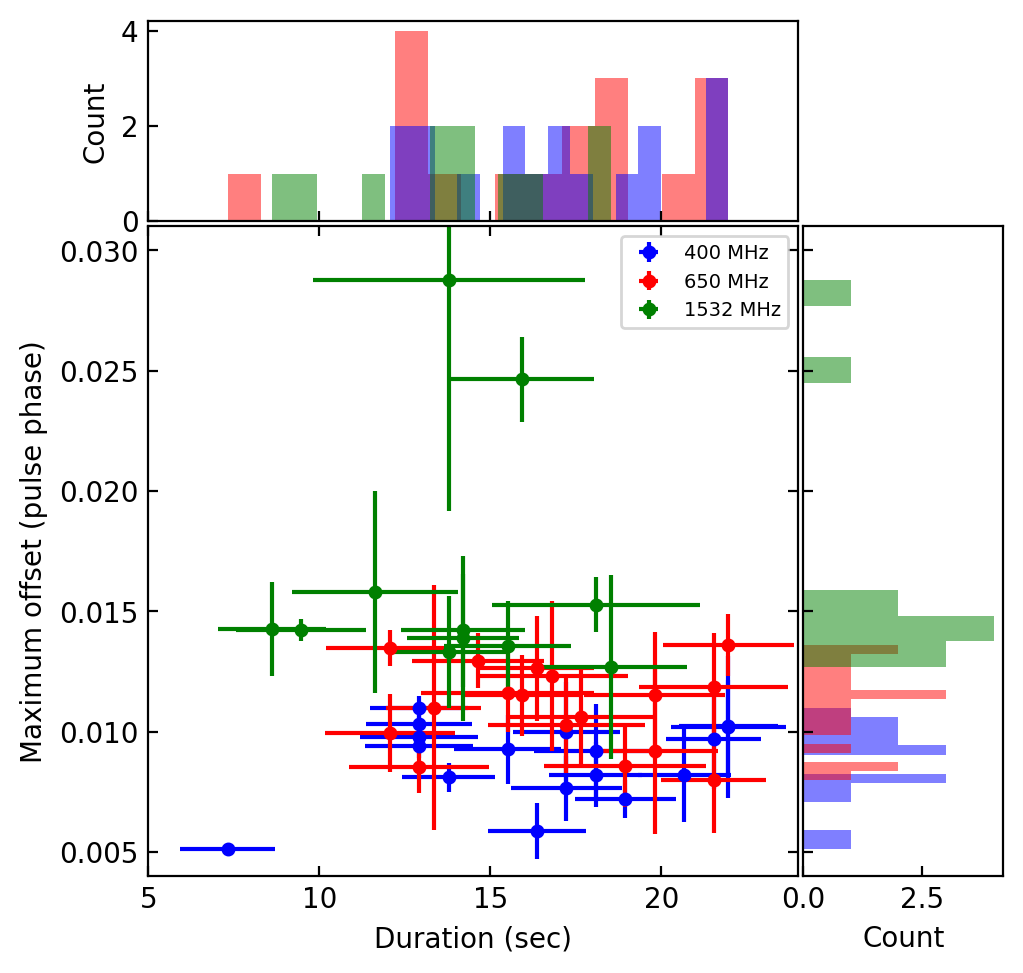}
\centering
\caption{
Duration of swooshes as a function of their maximum offset from the nominal fiducial phase for PSR~B0919+06 at multiple frequencies; 400 MHz GMRT ({\it blue}), 650~MHz GMRT ({\it red}), and 1532~MHz Lovell ({\it green}). See Section~\ref{b0919} for more details on the methodology to obtain this plot.
}
\label{fig:duroff}
\end{figure}


\begin{figure}
\centering
	\includegraphics[scale=0.32]{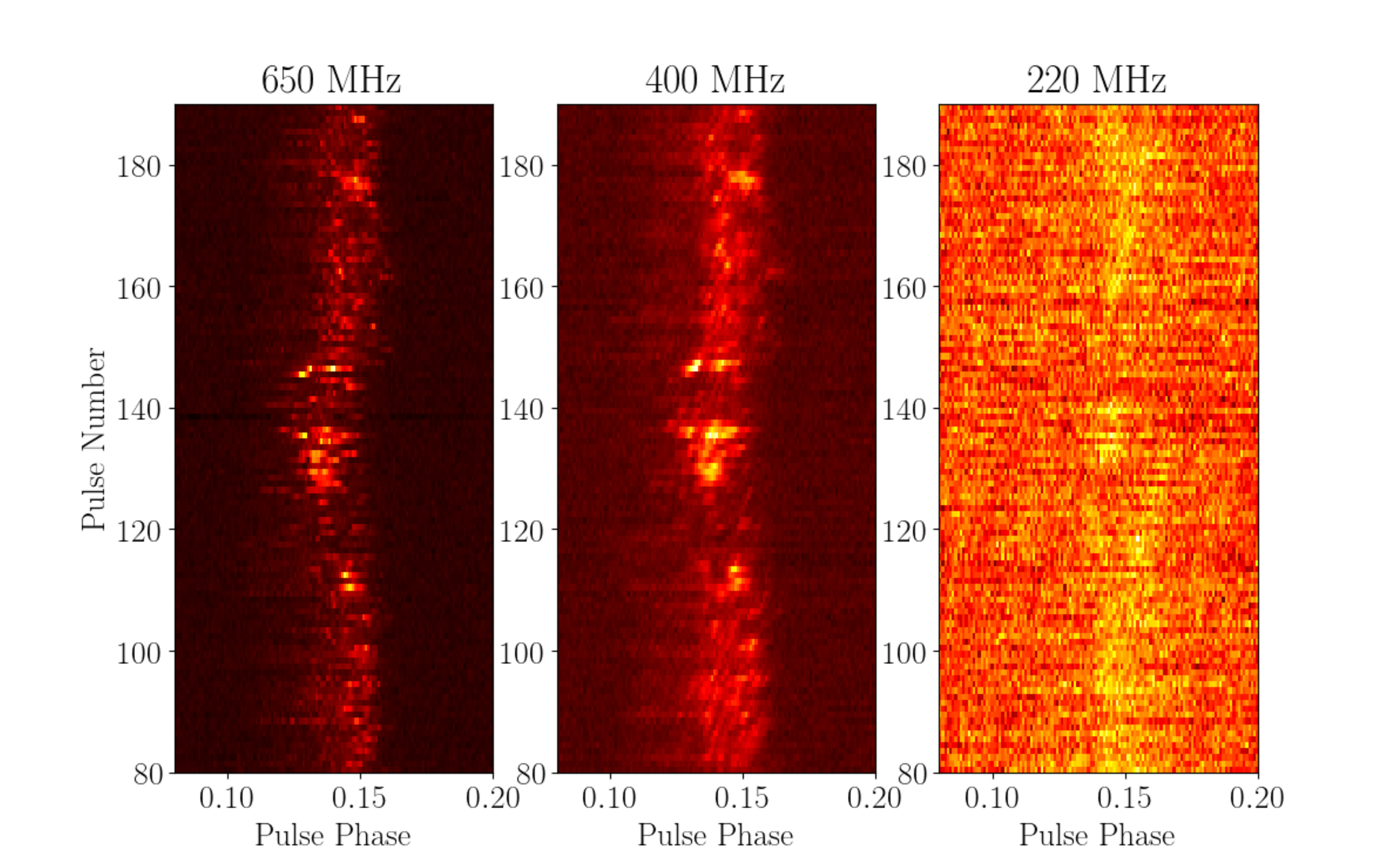}
    \caption{One swooshing event from PSR B0919+06 at three different radio frequencies. The bright pulses during the swoosh are clearly visible even at 220~MHz with no obvious offset from the normal emission while the rest of the pulses in the swoosh appear to be nulls.}
    \label{fig:b0919flare}
\end{figure}

\subsection{PSR B0919+06}
\label{b0919}
We performed a detailed analysis of the swooshes that were observed simultaneously at all the frequencies. Figure~\ref{fig:b0919_example} shows two examples of swooshes as seen at multiple frequencies. Firstly, we observe that not all swooshes show the same behaviour at lower radio frequencies. Some swooshes at high frequencies correspond to nulls at the lowest observed radio frequency while some swooshes show emission even at the lowest radio frequencies. Secondly, the magnitude of the phase offset during a swoosh is frequency dependent. In order to quantify this, we characterised each swoosh with a duration of the event and maximum offset measured in pulse phase. It is not trivial to model the shape of the swooshes with parameterized models as every swoosh looks very different in the phase versus pulse number space (see Figure~\ref{fig:b0919_example}). Moreover, measuring the duration of swooshes was challenging at lower frequencies as the jitter in pulse phase of single pulses is larger. Hence, we smoothed the pulse phases for each single pulse using a median filter with a kernel size that was optimally chosen to get minimal noise while still fully resolving each swoosh. The smoothed timeseries of pulse phases was then visually inspected to identify the duration of each swoosh. The maximum offset was then defined as the median of the five pulses around the pulse showing the maximum offset in phase. Figure~\ref{fig:duroff} shows the maximum of each swoosh and duration as a function of observing frequency, and it clearly shows the dependence of the swoosh amplitude with observing frequency such that the amplitude of the swoosh decreases with decreasing frequency.~\cite{shaifullah2017} had reported a peculiar behaviour of the swooshes at 150~MHz where they manifest themselves as partial or complete nulls. Since we observed simultaneously at a similar band (190--250~MHz) with the uGMRT, we were able to 
sample the swooshing behaviour at low frequencies. For most swooshes, we confirm the behaviour first reported in~\cite{shaifullah2017} where the swoosh at a higher frequency corresponds to either weak or a complete lack of radio emission at 220~MHz. Similar to~\cite{han2016}, we also looked at the distribution of the duration of swooshes for PSR~B0919+06 as a function of frequency and no correlation between the duration and observing frequency is observed. We do note that the estimation of the error on the duration was based on the assumption that the error on each phase measurement was small enough to result in a very small error on the estimated median during the smoothing process. Bootstrapping the pulse phase values near the visually estimated start point showed that the uncertainty introduced due to the median filtering would not be greater than 1 pulse. The error on the duration of the swoosh at a given frequency was scaled by the ratio of the variance of the pulse phase distribution at 1.4~GHz and the given frequency. We also acknowledge that the best way of getting a proper estimate of the error on the duration is by describing each swoosh with some analytical model which is beyond the scope of this paper.

\subsection{PSR B1859+07}

\begin{figure*}
\includegraphics[width=18cm]{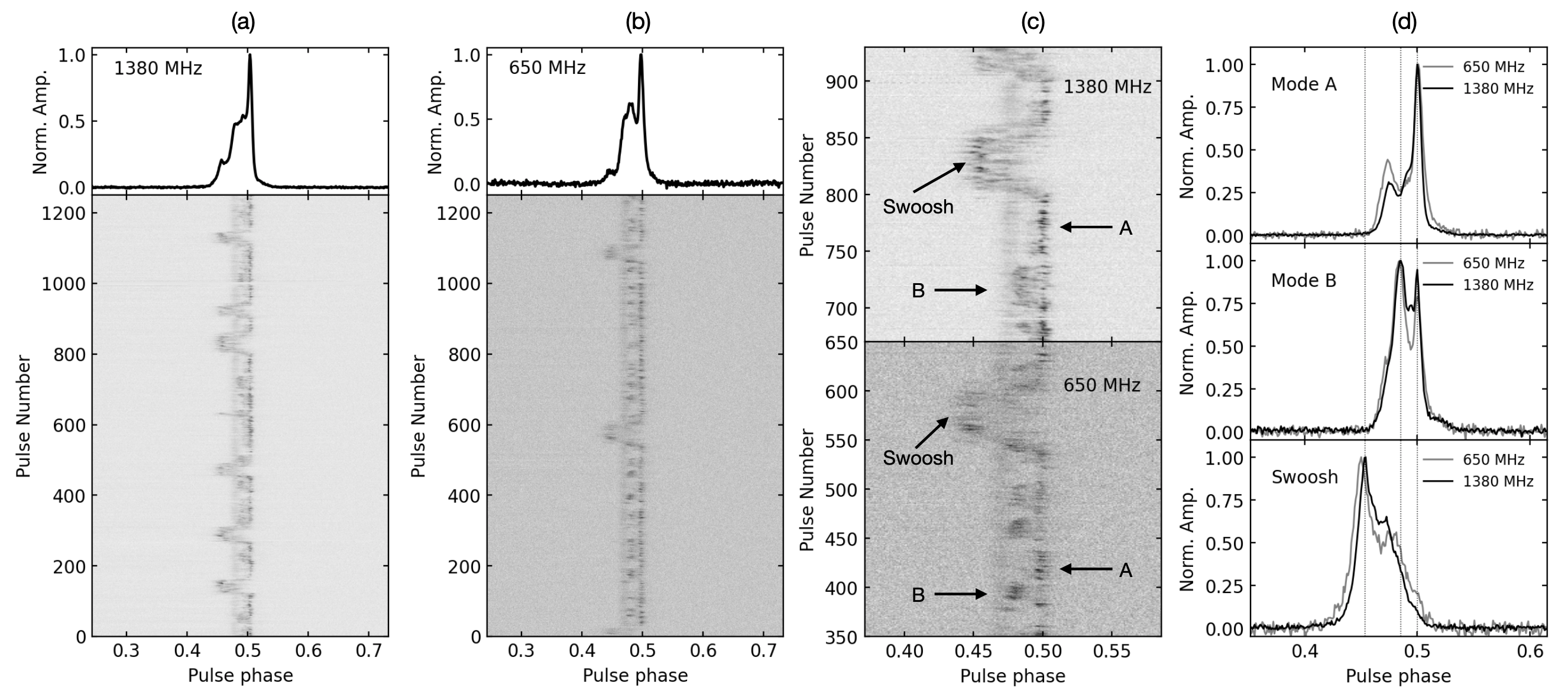}
\centering
\caption{
Non-simultaneous multi-frequency observations of PSR B1859$+$07. Panel (a) and (b) present 1250 single pulses of the pulsar obtained at 1380 and 650~MHz on MJD 58788 and 58455, respectively. The average pulse profiles are shown in the top panel. Zoomed versions of subsections of pulses are shown in panel (c) for both frequencies. In addition to the swoosh emission, the pulsar shows two different emission modes (A and B) within the normal emission state. The corresponding average pulse profiles from these three emission modes are compared in panel (d). The dotted vertical lines represent the peak pulse phases of the three modes at 1380~MHz.   
}
\label{B1859_sp}
\end{figure*}

The multi-frequency observations of PSR B1859$+$07 were processed in a similar fashion compared to the PSR~B0919+06 datasets. Figure~\ref{B1859_sp} presents the pulse sequences at 1380 and 650~MHz -- see panel (a) and (b) -- and both, swooshes and normal emission can be clearly seen. We note that the pulse sequence at 1380~MHz shows more swooshes compared to the low frequency. This is solely due to the swooshing rate of the pulsar at that particular epoch, and the 1380~MHz Arecibo and 650~MHz GMRT observations were not simultaneous. In addition to the swooshes, two distinct emission modes with different profile shapes are seen within the normal emission (see panel (c) in Figure~\ref{B1859_sp}). The switching between these two modes is abrupt and the emission is generally confined to the on-pulse region of the normal mode. 
The pulses of the three emission states were separated visually (avoiding transition pulses between states) and averaged together to form their pulse profiles (see panel (d) in Figure~\ref{B1859_sp}). As expected from radius to frequency mapping (RFM) in pulsars~\citep{cor78}, the pulse profile at higher frequency is narrower compared to that at lower frequency, and this is generally evident in all three emission states. 
In order to study the properties of the swooshes, we followed the same method described in section~\ref{b0919}. We note that the three emission states of the pulsar have different pulse profile shapes (see panel (d) in Figure~\ref{B1859_sp}). However, the pulse-to-pulse shape variations are large compared to the error on the phase offset that is computed by using a single template for all emission states in the cross-correlation process. Therefore, using a single template to obtain the phases of single pulses is accurate enough for our study. We also note that in general, the shape of the swooshes of PSR~B1859$+$07 is complex and longer in time compared to those of PSR~B0919$+$06. 
As before, we identified a swoosh as five or more consecutive pulses appearing earlier in pulse phase compared to the normal emission. The number of swooshes in our observations are given in Table~\ref{psr_info}.

\begin{figure*}
\includegraphics[width=17cm]{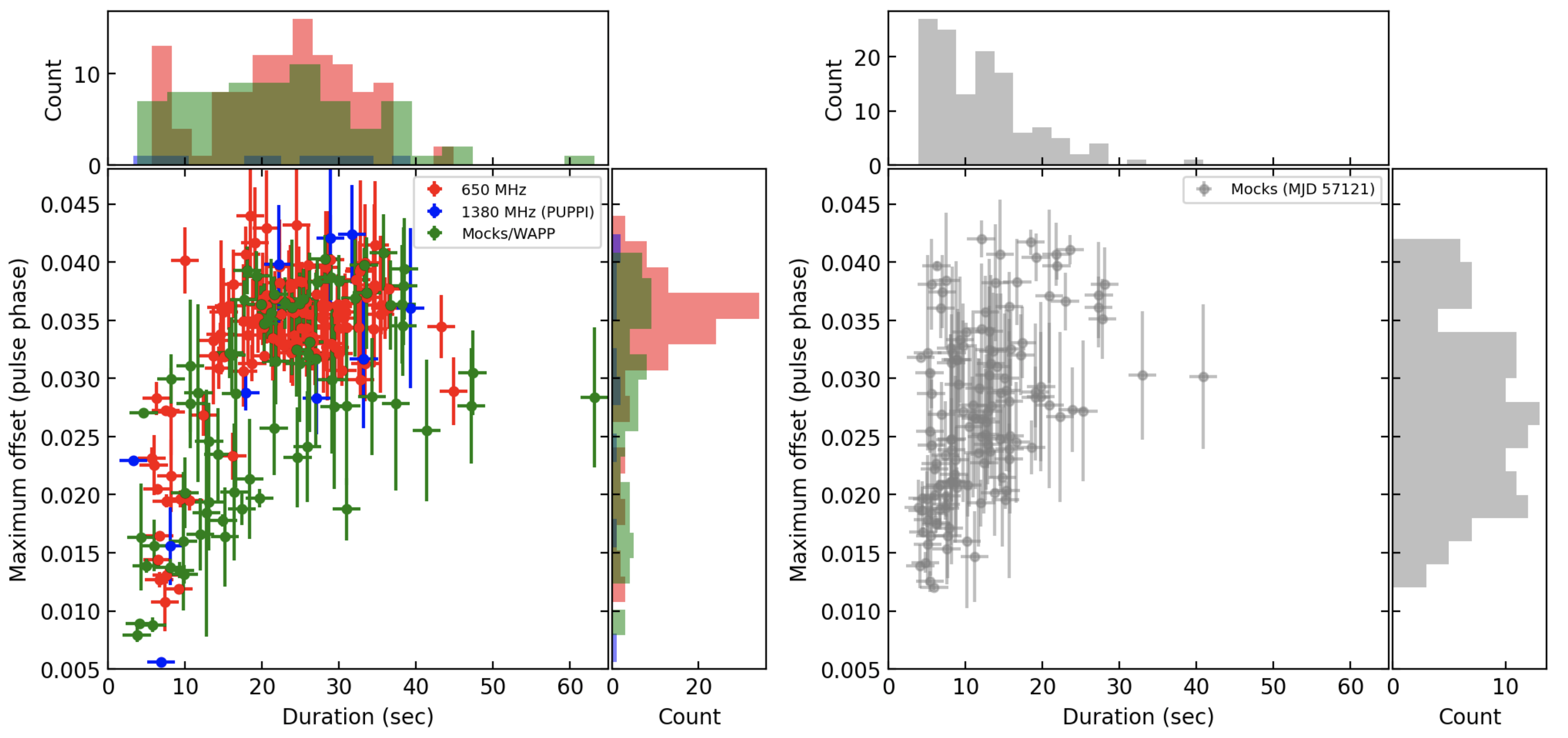}
\centering
\caption{
Same as Figure~\ref{fig:duroff}, but for PSR B1859$+$07. The {\it left} hand side panel shows the properties of swooshes obtained from all epochs, except MJD 57121. The {\it right} hand side panel shows the same properties obtained on MJD 57121. The different data sets are shown in different colors; 650 MHz GMRT ({\it red}), 1380 MHz Arecibo PUPPI ({\it blue}), Arecibo Mocks/WAPP data except on MJD 57121 ({\it green}), and Arecibo Mocks data on MJD 57121 ({\it gray}). The details of the observations can be found in Table~\ref{psr_info}. The swooshes in the Arecibo data on MJD 57121 show smaller duration in general compared to other epochs.}
\label{fig:B1859}
\end{figure*}

Following the same method in Sec~\ref{b0919}, we then estimated the properties of these swooshes for both frequencies (see Figure~\ref{fig:B1859}). 
We note that the swooshing rate of this pulsar is highly variable depending on the observation epoch. The GMRT observations were separated by just a few days and the swooshing rate ($\sim$0.4~min$^{-1}$) was consistent across each epoch.
However, there were long gaps between the Arecibo observations (five observations were taken over more than 16 years -- see Table~\ref{psr_info}) and the swooshing rate varied from epoch to epoch with the highest of 1~min$^{-1}$ on MJD 57121. 
We also note that most of the swooshes in the GMRT observations are concentrated around an offset of $\sim$0.035 in pulse phase. However, the Arecibo observations, except the one obtained on MJD 57121, show a more scattered distribution. 
Further, the duration of most swooshes seen on MJD 57121 is much smaller than the other epochs. 
This can be clearly seen in the duration distribution in Figure~\ref{fig:B1859}, which peaks around $\sim$6~s on MJD 57121 (i.e. the lower bound of the distribution). In contrast, the distribution of the duration on other epochs is approximately uniform. This suggests that the pulsar exhibits a variation in the swoosh properties and the swooshing rate over long intervals (i.e. on a month to year timescales).

\begin{figure}
\includegraphics[width=7.5cm]{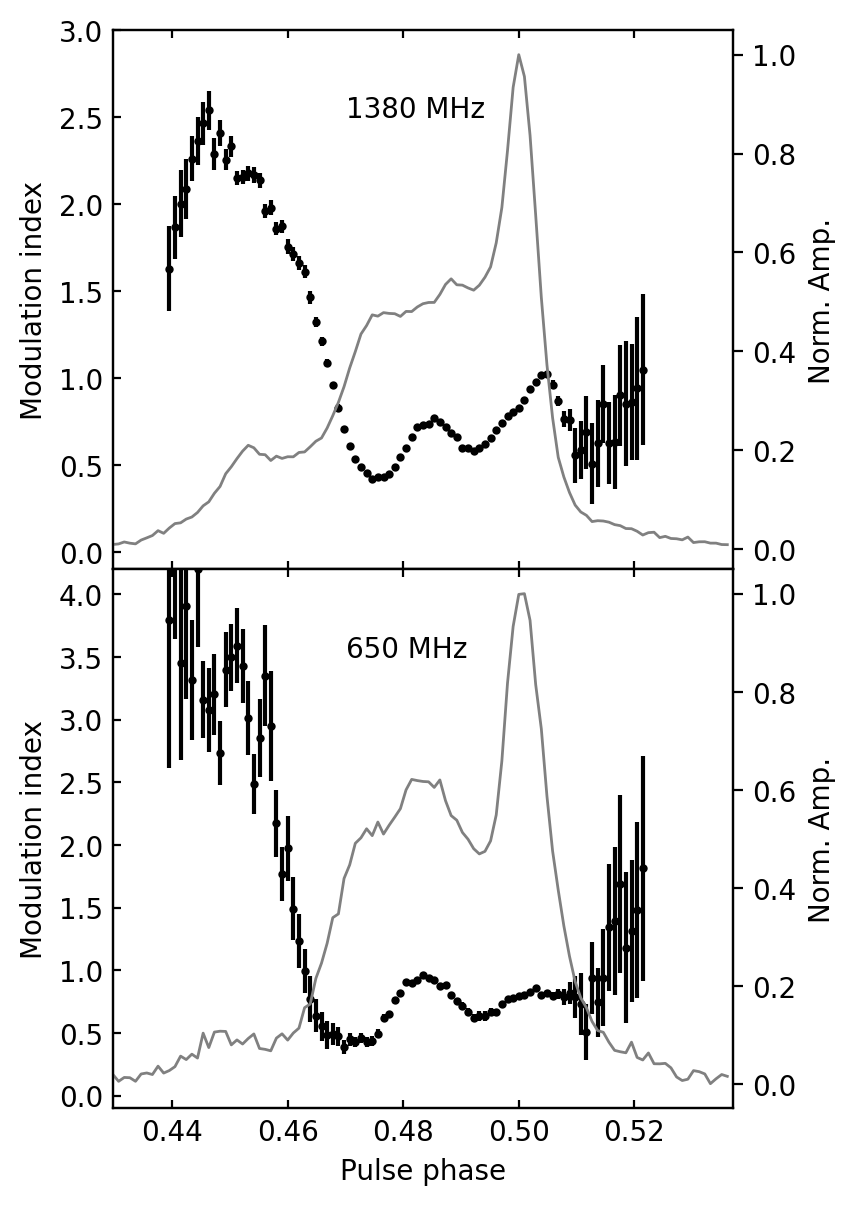}
\centering
\caption{
\textbf{Left Panel:} Modulation index for PSR B1859$+$07 using the pulse sequences shown in Figure~\ref{B1859_sp} at 1380 and 650~MHz. The average pulse profiles are over-plotted.    
}
\label{B1859_mod}
\end{figure}

\begin{figure*}
\includegraphics[scale=0.35,angle=-90]{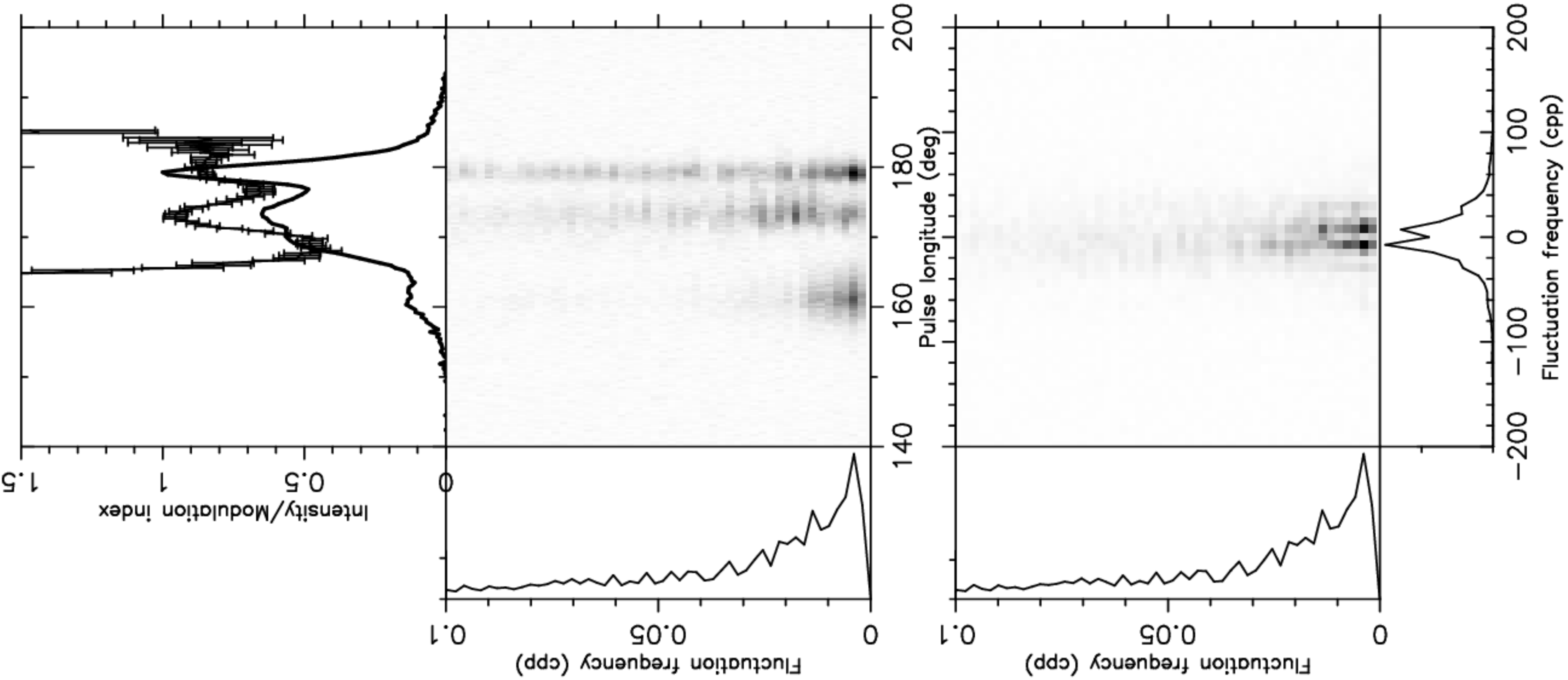}
\includegraphics[scale=0.35, angle=-90]{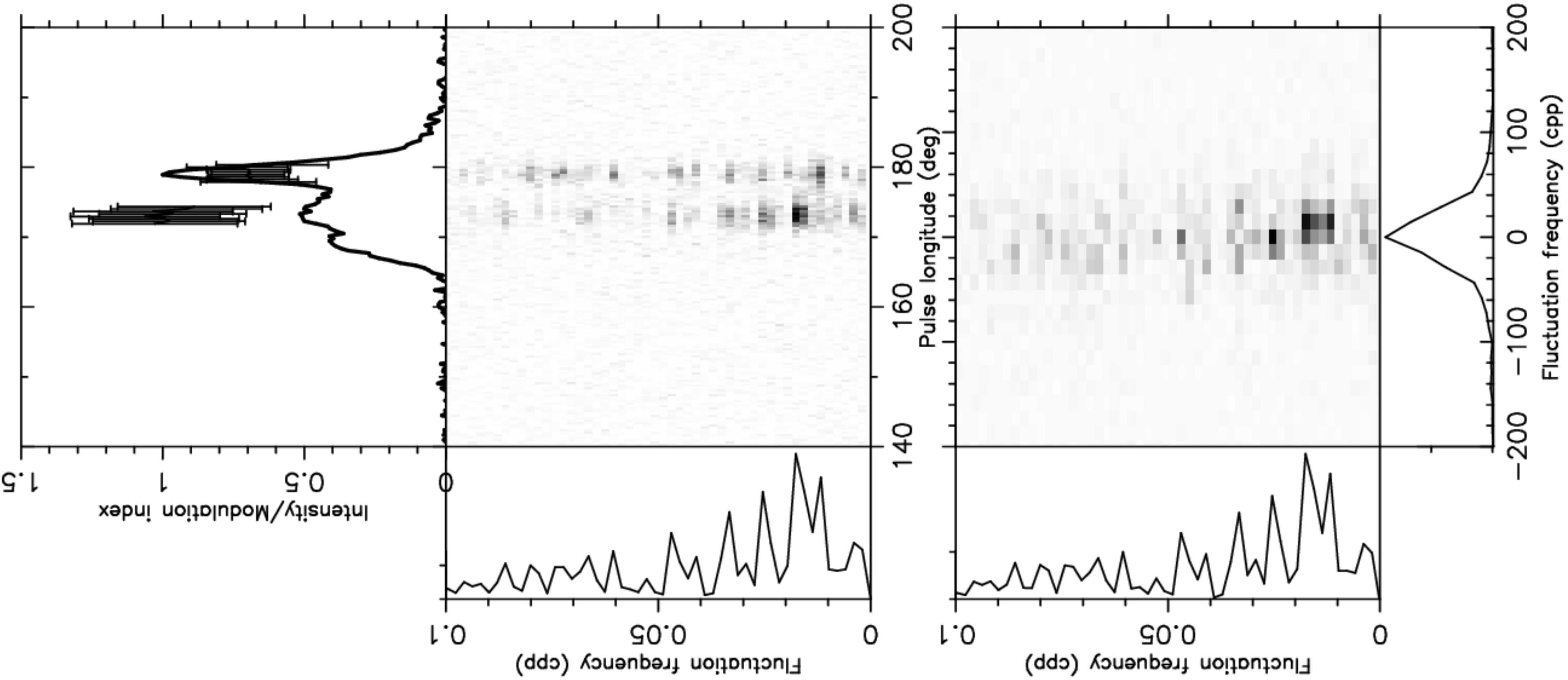}
\centering
\caption{\textbf{Left Panel:} Integrated pulse profile with the modulation index (Top -- similar to Figure~\ref{B1859_mod}), the longitude resolved fluctuation spectrum (middle) and 2-D fluctuation spectrum (bottom) of PSR B1859+07 at 650~MHz. \textbf{Right Panel:} The same plot for the same data set after removing the swooshes.}
\label{B1859_fs}
\end{figure*}

To further study the different emission modes in PSR~B1859+07, we estimated the modulation index across the radio emission phase. For a given pulse sequence at each frequency, the modulation index at a given pulse phase $i$ is calculated as  $m_{i}=\sigma_S/\langle S \rangle$, where $\sigma_S$ is the standard deviation of the fluxes along a $i$ and $\langle S\rangle>$ is their mean \citep[see][]{lk12}.  The results of this analysis are shown in Figure~\ref{B1859_mod}. The modulation is large around the leading edge of the pulse profile due to swooshes. A relatively large value of $m$ is also observed around the pulse phase 0.47--0.49 due to the switching between the A and B emission modes (see panel (c) in Figure~\ref{B1859_sp}).~\cite{wahl2016} showed that the swooshing in PSR~B1859+07 is quasi-periodic with a mean period of 150 rotation periods. In order to confirm this, we created a longitude resolved fluctuation spectrum (LRFS) and a 2-D fluctuation spectrum (2DFS) of the data at 650~MHz~\citep[see][and the references therein]{edwards2003} using \textsc{psrsalsa}~\citep{weltevrede2016}. Since some underlying modulation features may be hidden by the swooshes, we performed this analysis on both data sets including the swooshes and on the data after filtering out the swooshes (see Figure~\ref{B1859_fs}). In the entire data set, the LRFS is dominated by a low frequency peak around $\sim$0.006 cycles per period that is consistent with the period reported in~\cite{wahl2016}. The bi-modal structure in the 2DFS can be attributed to the ingress and egress of pulses at the beginning and the end of a swoosh. On the other hand, the LRFS of the data set after filtering out the swooshes shows Fourier power at around 0.02 cycles per period. The corresponding 2DFS for that feature is symmetric and does not show any skewness suggesting that the feature in the LRFS corresponds to a longitude stationary modulation. The feature possibly originates from the quasi-periodic switching between emission mode A and mode B as shown in Figure~\ref{B1859_sp}.

\section{Discussion}
\label{dis}

\subsection{Swooshes at multiple frequencies}

\begin{figure}
\centering
	\includegraphics[scale=0.28]{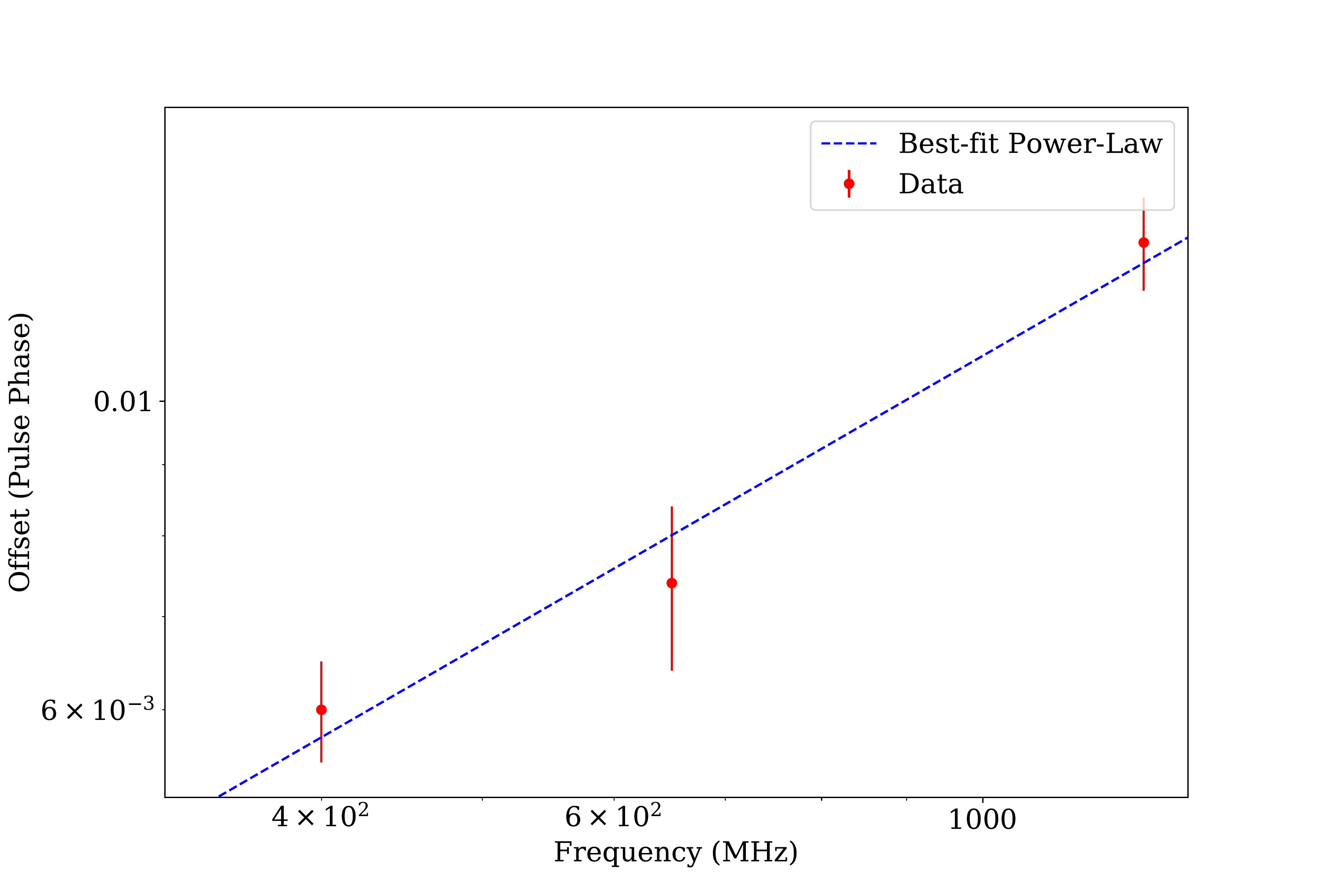}
    \caption{Offset from the nominal pulse phase at the peak of a swoosh as a function of observing frequency for PSR~B0919+06. The blue dashed line represents the best-fit power-law function.}
    \label{fig:b0919flarefit}
\end{figure}

\begin{figure*}
	\includegraphics[scale = 0.45]{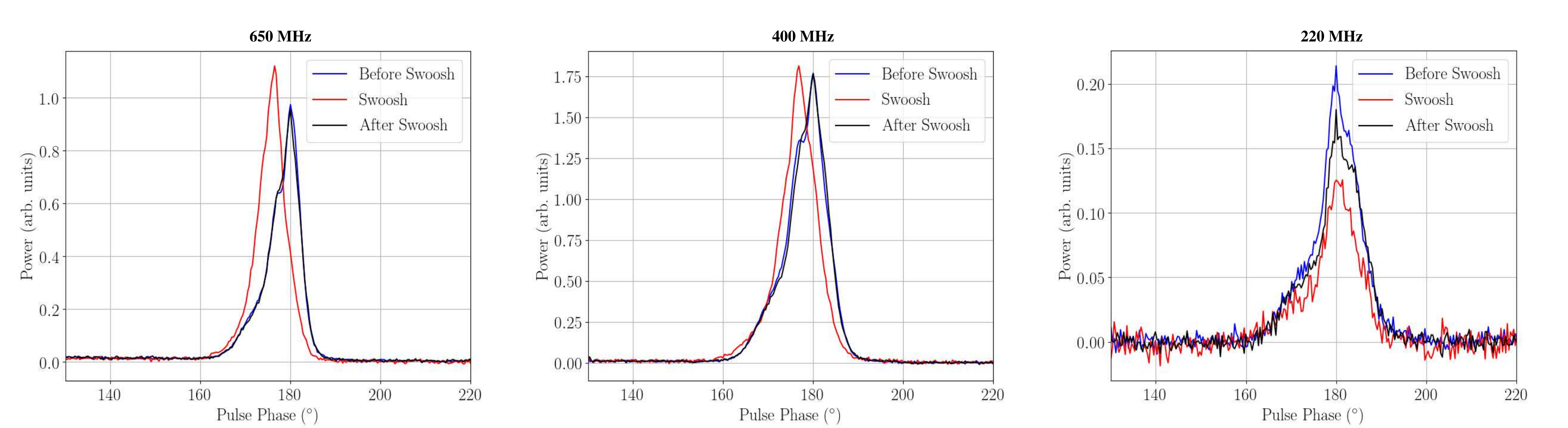}
    \caption{Folded pulse profile of PSR~B0919+06 at different observing frequencies. For each frequency, the blue profile is the averaged profile before a swoosh like event, the red profile is during a swoosh and the black profile is after a swoosh.}
    \label{fig:0919allprofs}
\end{figure*}

From the 17 swooshes that were identified for PSR B0919+06, 11 swooshes were identified at all three bands of GMRT and at 1.4~GHz from Lovell data. In order to quantify the relationship between the phase offset during a swoosh and the observing frequency, we characterised the phase offset with a power-law function such that,
\begin{equation}
    \phi(\nu) = k\,\nu^{-\alpha},
    \label{eq:pl}
\end{equation}
where $\phi(\nu)$ is the maximum phase offset from the nominal emission phase during a swoosh, $\nu$ is the observing frequency and $\alpha$ is the power-law index. We note that we only used the power-law relationship for a qualitative analysis of the relationship between swooshes and the observing frequency and it was not physically motivated. We fit all 11 swooshes of PSR~B0919+06 using the equation~\ref{eq:pl}. Figure~\ref{fig:b0919flarefit} shows one such swoosh with the best fit power-law function. On average, the best-fit value of $\alpha$ for all swooshes ranged from 0.35--0.6 with a mean value of 0.51$\pm 0.3$ suggesting that there is a consistent relationship between the emission mechanism causing the swoosh and the emitted frequency. Previously,~\cite{shaifullah2017} and~\cite{yu2019}, have shown that pulses during a swoosh appear to null at lower frequencies. We note that in our dataset, while that behaviour is generally consistent with the conclusions by previous authors, we see a varied behaviour in radio emission at 220~MHz during a swoosh. For example, there were swooshes where no change in radio flux or phase offset is observed at 220~MHz (top panel of Figure~\ref{fig:b0919_example}) and in some cases, nulling is only observed in a fraction of pulses in a swoosh at 220~MHz (see Figure~\ref{fig:b0919flare}) where the brightest pulses during the swoosh are easily visible while others appear as nulls. In order to study the radio emission during the swoosh, we created an average pulse profile for B0919+06 by adding pulses just before a swoosh, during a swoosh and right after a swoosh for all 9 swooshes at different frequencies and the results are shown in Figure~\ref{fig:0919allprofs}. A couple of things stand out in these profiles: most importantly, the ratio of the amplitude of the profile during a swoosh to the amplitude before and after the swoosh keeps decreasing with observing frequency. At higher frequencies, pulses during a swoosh appear to be brighter and are progressively dimmer at lower and lower frequencies. Based on these observations, we propose that at lower frequencies, the pulsar never really stops emitting during a swoosh but rather the pulses are faint enough that they are below the detection threshold of our telescopes and appear as nulls. 

Below, we discuss our observations and results in the context of emission mechanisms that have been proposed over the course of the last two decades. Swooshes of PSR B0919+06 have been studied in detail in the past~\citep{rankin2006, perera2015, han2016}. In these studies, the authors had already showed the bi-modal nature of emission in PSR~B0919+06 and multiple models were proposed to explain them. The absorption model was proposed to explain the complete lack emission at lower frequencies~\citep{rankin2006}. Quasi-periodic swooshing behaviour is evident in PSR~B1859+07 while it is clearly sporadic in PSR~B0919+06~\citep{han2016}. Orbital modulation has been proposed as an explanation for the quasi-periodic swooshes before~\citep{wahl2016}. If we assume that both pulsars share a similar emission mechanism, it is highly unlikely that a binary orbital modulation is responsible for the swooshing behaviour. One can still get sporadic swooshes if they can be attributed to interaction of the pulsar with a debris disk around it. Such debris disks are postulated to be formed from fall-back of the material ejected during the Supernova explosion that created the neutron star. Such a model was proposed to explain spin-down and profile variations in PSR~J0738$-$4042~\citep{brook2014}. For this model to work, an interaction of debris with the neutron star magnetosphere should cause a change in the magnetospheric currents and thus the net torque on the neutron star. This means that the spin-down in a pulsar should be correlated with swooshes. No such correlation has been seen for either PSR~B0919+06 or PSR~B1859+07 \citep[see][]{perera2015,perera2016}. It is important to note that when these models were proposed, the observations mainly lacked simultaneous observations over a wide range of frequencies. Recently, multi-frequency studies were performed by~\cite{shaifullah2017} and \cite{yu2019} where the chromatic behaviour of swooshes was seen for the first time. They clearly show that if absorption is to explain the emission in PSR~B0919+06, it has to be extremely complicated and variable in time and observing frequency and hence, unlikely to be the case here. Moreover,~\cite{yu2019} found some peculiar emission properties of swooshes that were studied over a large (0.27--1.6~GHz) frequency range. In particular, they found that not all swooshes at higher frequencies are associated with pseudo-nulls and the folded profile during the swoosh showed evidence for a different spectral index. Our detailed analysis in this observing campaign confirms these findings by previous authors.

\subsection{Swooshes in context of the core-cone model}

There are several beam models put forward to explain the observed emission of pulsars over the last 50 years since their discovery~\citep[see][for a review]{melrose2017}. 
The core-cone model has been used extensively to explain pulse profiles with different emission components \citep[see][for an extensive review]{rankin83a,rankin83b,rankin90,rankin93} over several decades. Previous polarimetric studies of PSR~B0919+06 and PSR~B1859+07~\citep{rankin2006} posit that the pulse profiles are consistent with a triple profile with a core component in the central emission region surrounded by a conal component pair of emission along the periphery of the emitting region. 
~\cite{rankin2006} suggest that the radio emission in these pulsars can be attributed to a partially active cone -- the leading part of the cone is inactive during the normal emission and the trailing part during the swoosh emission. The core emission can be weak or completely inactive during the swoosh, depending on the strength of the event (or the offset of the swoosh from the normal emission)~\citep[see][for a detailed review]{rankin2006b}. 
The swoosh will be seen differently at different frequencies as the radius of the cone will change at different emission heights. 
In this model, the drifting sub-pulses result from emitting `beamlets' or `sparks' rotating due to the $\bar{E}{\times}\bar{B}$ drift ~\citep{deshpande2001,herfindal2007,force2010}. 
Since, PSR~B0919+06 and PSR~B1859+07 do not exhibit drifting sub-pulses, one can suppose the swooshes are due to the motion of a single rotating spark within the conal beam. The onset of the swoosh is seen as the spark rotates away from the observer and the gradual return is seen when the spark returns to the nominal phase after completion of one full rotation within the carousel. 


Although one can attempt to explain the swooshes in the core-cone framework, there are some major caveats that may render this model unsuitable. Firstly, swooshes seen in these pulsar show a large diversity in their shapes and are highly asymmetric. The pulses during the swoosh tend to stay at an earlier phase for few tens of rotations before coming back to the nominal phase. These observations cannot be explained by a single spark rotating in the beam carousel as one would always expect a symmetric swoosh with no more than a couple of pulses at the maximum offset during the swoosh.  Secondly, based on RFM~\citep{cor78}, one would expect the maximum offset during a swoosh to be lower at higher frequencies as the radius of the conal component is smaller. The observations show exactly the opposite behaviour from what is expected from the core-cone model with lower offsets at lower radio frequencies (Figure~\ref{fig:duroff}).~\cite{yuen2017} were able to reproduce the swooshes in PSR~B0919+06 by assuming a slightly modified version of the core-cone model where the drift rate of the beamlet changes due to a global change in the magnetosphere. While such a change in the drift rate can manifest itself as a swoosh, a chromatic dependence of the offset suggests a dependence of the drift rate on the height of radio emission which this model does not take into account.

\subsection{Alternate Emission Model}

\begin{figure*}
\includegraphics[width=16cm]{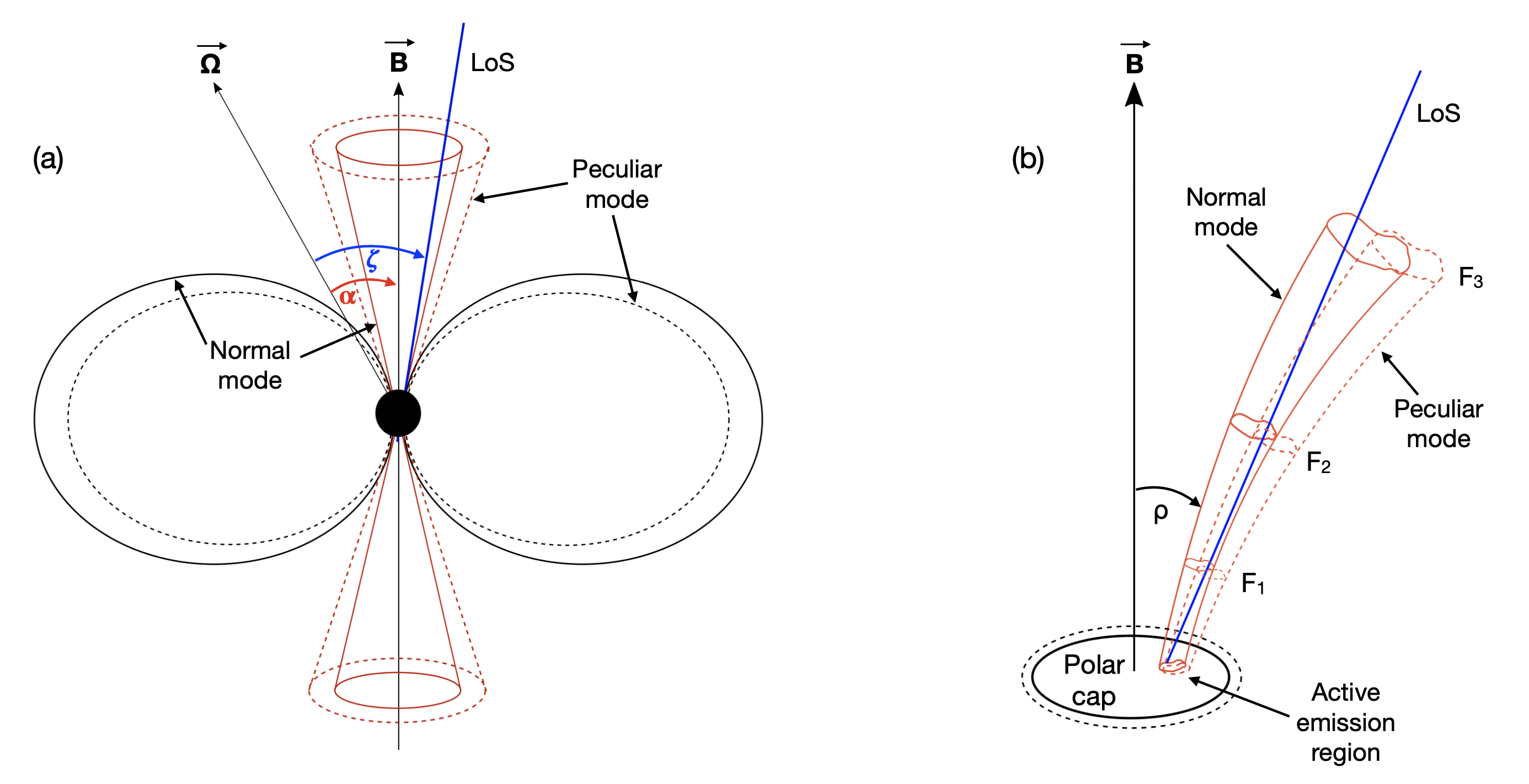}
\centering
\caption{
The schematic diagram of the emission model. The rotational and magnetic axes are denoted by $\vec{\Omega}$ and $\vec{B}$, respectively. The magnetic axis is misaligned by an angle $\alpha$ with respect to the rotation axis and $\zeta$ is the view angle of the LoS with respect to the rotation axis. (a) The closed field lines of the dipolar magnetosphere ({\it black}) and the emission beams ({\it red}) are presented. The configuration for the regular emission mode is presented in solid lines and the peculiar emission mode is presented in dashed lines. During the peculiar emission, the magnetosphere shrinks and the emission cone gets expanded, and the LoS cuts a different emission region across the beam. (b) Only a patch of the polar cap region is assumed to be active. The frequency-dependent emission produces from this active region and the flux tubes are presented for the normal mode ({\it red solid} lines) and the peculiar mode ({\it red dashed} lines). Three frequency-dependent emission generation regions are denoted ($F_1$, $F_2$, and $F_3$) with frequencies $F_1 > F_2 > F_3$. The opening angle of the emission flux tube from the magnetic axis is denoted as $\rho$. According to this particular schematic view, the emission can be detected at frequencies $F_1$ and $F_2$ during both emission modes, but not at $F_3$ during the peculiar mode as its flux tube is out of our LoS. This explains the observed emission nulling at low frequencies for PSR B0919$+$06. 
}
\label{diagram}
\end{figure*}

Below, we describe a simple emission model that can explain most of the key emission features of the pulsars presented in this paper. It draws significantly from the model proposed in~\citet{tim10} that explains nulling and mode changing in pulsars by shrinking and expanding the magnetosphere of the neutron star.
The model assumes a dipole magnetic field line structure and a schematic diagram is shown in Figure~\ref{diagram}. The emission beam is assumed to be a `fan beam' with multiple flux tubes along the open field lines~\citep{karastergiou2007,oswald2019}. We assume the fan beam to be ``patchy'' and partially filled meaning that only a portion of the tube is emitting radiation. The patchy beam scenario is one of the common beam models that has been used to explain the emission properties of pulsars \citep[see][]{lyne1988,manchester2010}. 
 As shown in Figure~\ref{diagram}, it is assumed that the emission patch is fixed within the polar cap region; i.e., the region consists of open magnetic field lines with respect to the boundary of the comoving magnetosphere (known as the light cylinder). When the magnetosphere shrinks the polar cap region expands accordingly with the dipole field lines and changes the orientation of the emission flux tube produced from the patchy region (see Figure~\ref{diagram} (b)). Due to these changes, our Line of Sight (LoS) encounters a different part of the flux tube, resulting in variation in the observed pulse profile shape, switching the pulsar from the normal to swoosh emission. 
It has been postulated that low frequency radio emission is produced at a higher altitude compared to the high frequency emission in the magnetosphere \citep[ie -- RFM][]{cor78, oswald2019}. Therefore, due to the curvature of the dipole magnetic field lines, the flux tube of the patchy beam at a lower frequency can be oriented in a direction that is out of our LoS (see Figure~\ref{diagram} (b)). This results in apparent nulls at low frequencies, which is consistent with the observations of PSR B0919$+$06 \citep[see Figure~\ref{fig:b0919_example}, and also][]{shaifullah2017}.

\subsubsection{Application to PSR B0910+06 and PSR B1859+07}
\begin{figure}
\includegraphics[width=8cm]{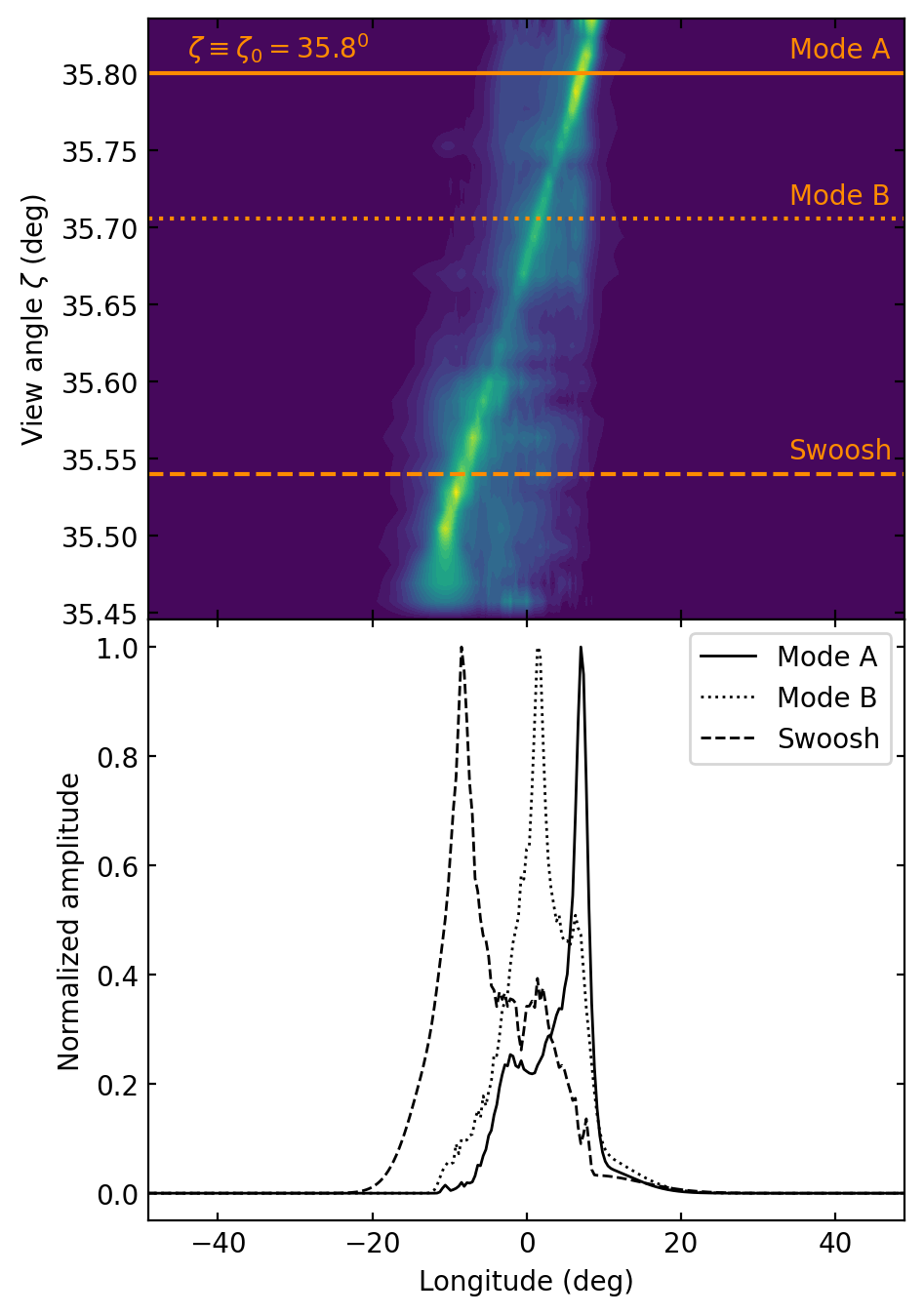}
\centering
\caption{
The sky map of the projected emission beam patch of PSR B1859$+$07 at 1380~MHz ({\it top}). The y-axis is the viewing angle $\zeta$ with respect to the pulsar rotation axis. The viewing angle of the LoS is $\zeta_0 = 35.8^\circ$. During the shrinkage of the magnetosphere, the projected emission patch moves to higher $\zeta$ values. The horizontal lines represent the trajectories for different emission modes across the beam. A cross section of the map gives the pulse profile of that particular viewing angle. The synthesised pulse profiles for each emission mode are given in the {\it bottom} panel. These profiles are largely consistent with the observed profiles for this pulsar (see Figure~\ref{B1859_sp} (d)).
}
\label{B1859_map}
\end{figure}

\begin{figure*}
\includegraphics[width=17cm]{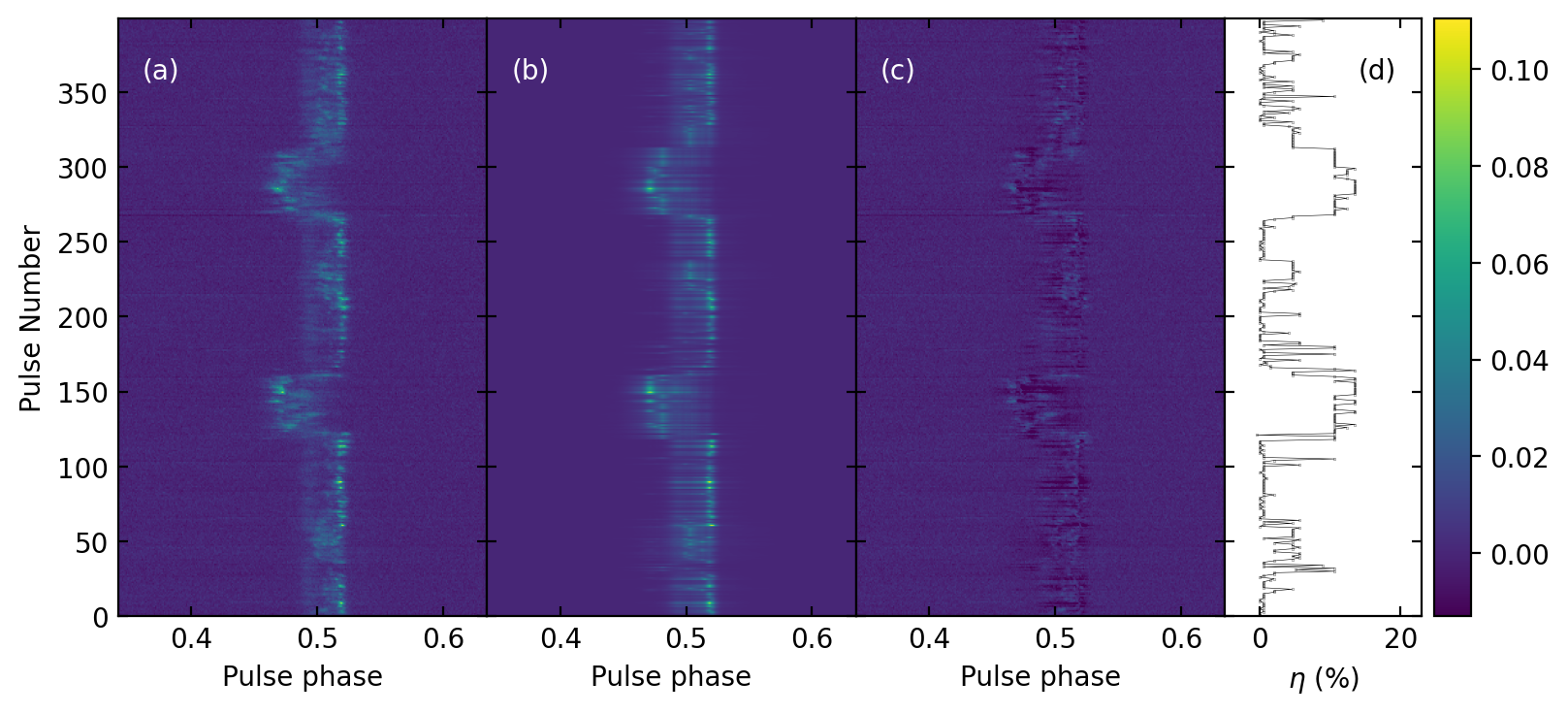}
\centering
\caption{
Comparison between the observed and simulated pulses of B1859$+$07 from the beam model. (a) First 400 observed pulses shown in Figure~\ref{B1859_sp} at 1380~MHz, including two swooshes; (b)Simulated pulses from the emission model; (c) Residuals after subtracting the simulated pulses from the observed pulses shown in (a); (d) Percentage of magnetosphere shrinkage required to produce the simulated pulses shown in panel(b).
}
\label{B1859_sim}
\end{figure*}

Using the above beam model, we then synthesised the emission beam in order to explain the pulse profiles of our pulsars. We used the pulse sequence of PSR B1859$+$07 at 1.4~GHz shown in Figure~\ref{B1859_sp}. We fit several Gaussians to each individual pulse in the pulse stack to generate noise free individual profiles. We then split the on-pulse region (including both normal and swoosh emission) into segments with sizes of two phase bins. We obtained the phase of pulse peak of each individual noise-free profile and grouped them into on-pulse segments according to their peak phases. Each profile group was then averaged to obtain a high S/N longitude-resolved profile. These grouped profiles represent all the emission features, including normal emission, swoosh emission, and transition emission between these two modes. We finally collated these profiles to create the sky-map of the beam (i.e., the emission beam projection on the sky when the pulsar is rotating) to satisfy the proposed emission beam model -- i.e., our LoS crosses the normal emission region of the beam when the magnetosphere is at its standard configuration, while the LoS crosses the swoosh emission region when the magnetosphere shrinks.
Figure~\ref{B1859_map} shows the sky-map of the beam and the y-axis is the viewing angle $\zeta$ with respect to the pulsar rotation axis (see Figure~\ref{diagram}), and the x-axis is the rotational longitude of the pulsar. 
Simply, a cross-section of the sky-map at a given $\zeta$ gives the pulse profile at that particular viewing angle.
The sky-maps are generally formed in pulsar magnetosphere modelling by tracing the photons from magnetic field lines and projecting them on the sky to explain the observed emission features \citep[e.g.,][]{dyks2003,dyks2004,perera2014}. In this paper, we did not attempt to model the magnetosphere as it is beyond the scope of this paper, rather we focused on the dipole field line structure and created the sky-map based on the above approach. We then linked the resultant sky-map to the expanding-shrinking magnetosphere with the known geometry of the pulsar to explain the observations.
Based on the polarimetric measurement, \citet{rankin2006} reported the geometry of PSR B1859$+$07 with a magnetic inclination angle $\alpha \approx 31^\circ$ and a LoS view angle $\zeta_0 = \alpha+\beta \approx 35.8^\circ$, where $\beta$ is the closest approach of the magnetic axis to the LoS. 
The opening angle $\rho$ of the emission patch with respect to the magnetic axis can be written as $\rho \approx (3/2)\theta_{\rm em}$, where $\theta_{\rm em}$ is the colatitude of the emission \citep{gangadhara2001}. For polar cap emission, the opening angle can be approximated as $\rho \propto \sqrt{r_{em}/R}$, where $r_{em}$ is the emission altitude and $R$ is the size of the magnetosphere \citep[see Section 3.3 and 3.4 in][for further details]{lk12}. If the magnetosphere is shrunk by a small fraction of $\eta$, then the polar cap region expands and thus, the emission patch moves to a larger opening angle. The new opening angle can be written as $\rho = \rho_0\sqrt{1/(1-\eta)}$ for a fixed emission altitude, where $\rho_0$ is the opening angle when the magnetosphere is at the normal emission. 
In the projected sky map, this magnetosphere expansion  can be viewed as moving the emission patch to higher viewing angles, resulting in that the LoS ($\zeta_0 = 35.8^\circ$) encounters different parts of the beam, producing different emission features and pulse profile shapes. 
A 10\% shrinkage of the magnetosphere from the normal configuration moves the emission patch by $\sim 0.3^\circ$ to higher $\zeta$ values, allowing our LoS to cross the swoosh emission part of the beam (see Figure~\ref{B1859_map}).
Note that the y-axis in Figure~\ref{B1859_map} is scaled according to the shrinkage percentage we used in the model and therefore, it is rather arbitrary.
By a $\sim$5\% shrinkage from the normal mode, we can observe the emission from Mode B of the pulsar. The bottom panel of Figure~\ref{B1859_map} shows the synthesised pulse profiles based on our simulation with shrinkage of 0, 5, and 10\% for normal (Mode A), Mode B, and swoosh emission, respectively. These model-predicted profiles are largely consistent with the observed profiles shown in Figure~\ref{B1859_sp}. In order to show how well the model can explain the observations, we simulated the first 400 pulses in Figure~\ref{B1859_sp} observed at 1380~MHz. Figure~\ref{B1859_sim} shows our results and the residuals between the observed and the simulated pulses indicate that the model is capable of explaining the observations well. The last panel in the figure presents the degree of shrinkage of the magnetosphere that is required to produce the simulated pulses shown in panel (b). In general, the swoosh emission can be explained with a shrinkage of approximately 10\%, but there are a few pulses within the swoosh emission that require a greater shrinkage ($\sim$14\%) to produce the observed emission.
Similarly, we can explain the observed swooshes of the pulsar at 650~MHz.

\begin{figure}
\includegraphics[width=8cm]{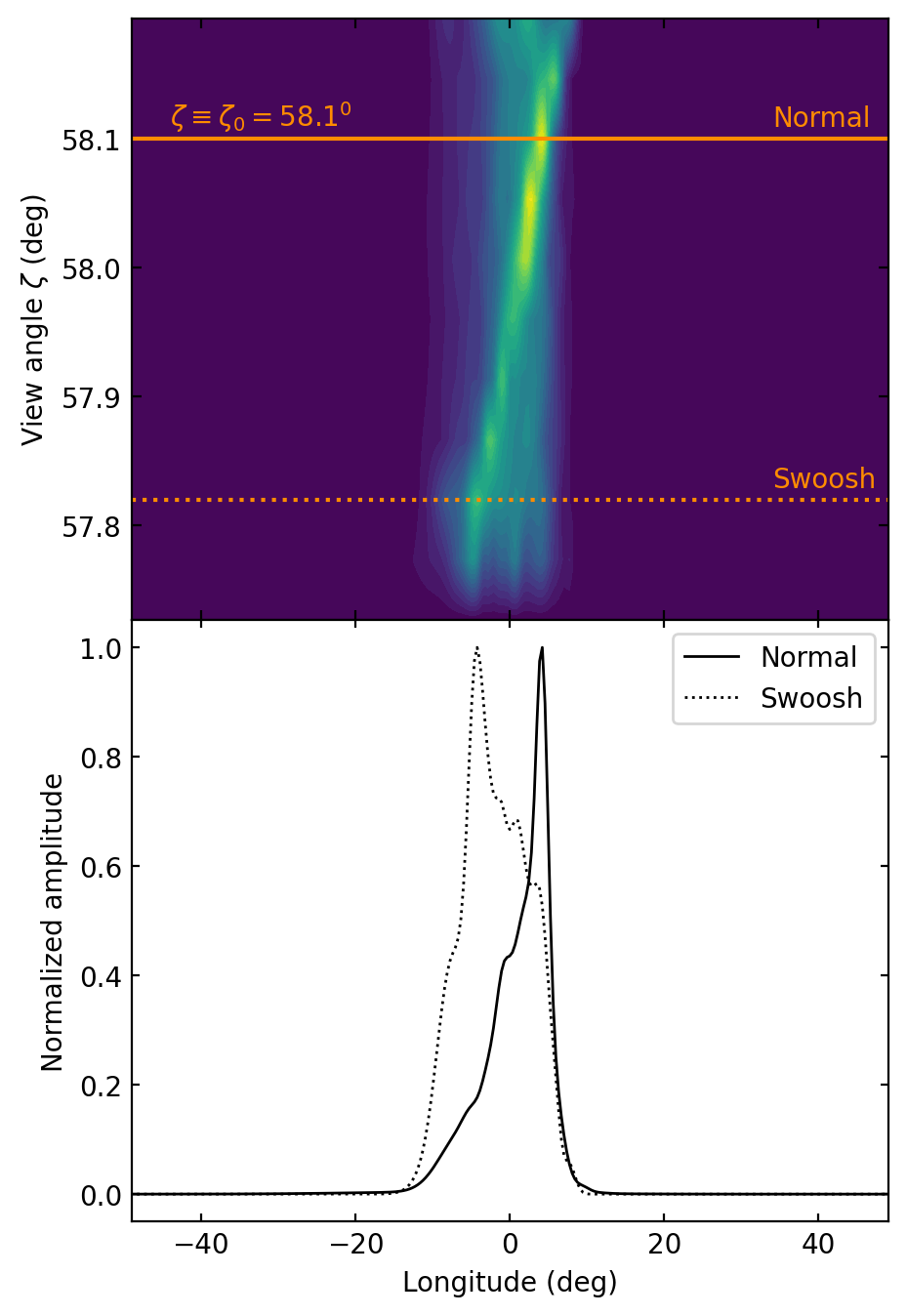}
\centering
\caption{
Same as Figure~\ref{B1859_map}, but for PSR B0919$+$06 at 650~MHz. The viewing angle of our LoS is $\zeta_0 = 58.1^\circ$.
}
\label{B0919_map}
\end{figure}

Following the same procedure as described above, we obtained the sky-map of the emission beam of PSR B0919$+$06. We note that our 1380~MHz Lovell telescope observations of the pulsar showed a significant amount of RFI. Thus, we used 650~MHz GMRT data to map the emission beam. The sky-map of the beam is shown in Figure~\ref{B0919_map}. The polarimetry of the pulsar constrained its geometry to be $\alpha \approx 53^\circ$ and $\zeta \approx 58.1^\circ$ \citep[see][]{rankin2006}. Assuming 10\% shrinkage of the magnetosphere during the normal to swoosh emission, as before for PSR B1859$+$07, we estimated that the emission patch moves by $\sim0.6^\circ$ on the sky to higher view angles. Then, the swoosh emission of the beam aligns with the view angle of our LoS. The bottom panel in Figure~\ref{B0919_map} again shows the simulated pulse profiles for the normal and swoosh emission, which are largely consistent with the observed profiles.

As shown in Figure~\ref{fig:b0919_example}, PSR B0919$+$06 exhibits quasi-nulling at low frequency emission at 220~MHz. As mentioned before, the low frequency emission is created at high altitudes in the magnetosphere according to RFM. Thus, the observed quasi-nulling can also be explained with our simple emission beam model as the shrinkage of the magnetosphere during the swoosh moves the high altitude flux tube outside our LoS (see Figure~\ref{diagram} (b) for details). The nulling cannot be seen all the time in our observations at low frequency and this is driven by the amount of shrinkage of the magnetosphere. We also note that the bright correlated emission at all frequencies during a swoosh cannot be explained with our simple beam model (Figure~\ref{fig:b0919flare}). This will require further development of the magnetospheric model to explain these multi-frequency correlated features in swooshes and is beyond the scope of this work.

As described in \citet{tim10}, the dipole magnetosphere with the force-free condition is capable of explaining a set of metastable states (i.e. magnetosphere configurations that can persist for a much longer time than the period of the pulsar) with changes in the magnetosphere size. According to the model, more open field lines are accompanied with a smaller magnetosphere and a larger beam during the swoosh emission and thus, the spin-down switches to a high state, releasing more energy compared to the normal emission mode. The previous studies show that these two pulsars exhibit long-term quasi-periodic spin-down variations ($\sim$order of hundreds of days), mainly switching between two spin-down states  \citep[see][]{perera2015,perera2016}. However, they could not find a conclusive evidence for a correlation between these long-term spin-down state variations and short-term swooshes ($\sim$ order of seconds -- see Figures~\ref{fig:duroff} and \ref{fig:B1859}). The physical mechanism behind switching between metastable states is not clear; however, \citet{tim10} discussed that it could be due to a combination of different current density distributions and different sizes of the corotating magnetosphere. 
As we showed in our simple model, the magnetosphere should consist of a short-term metastable state in order to explain the swoosh emission.  
The same mechanism could be responsible for these short-term states, but a further investigation is necessary to understand the physics behind these short-term metastable states and the mechanism that drives the change between different metastable states. This investigation is out of the scope of this work.

 The model described above can be invoked to explain the fast mode changes in PSR~B0611+22 where an offset in pulse phase is observed every 2500 rotations~\citep[see][for more details]{rajwade2016}. One caveat we would like to point out is that the mode change in PSR~B0611+22 is much faster ($\sim$1-2 pulse periods) than the time it takes to reach maximum phase offset during a swoosh. Regardless, they can be attributed to local changes in the magnetosphere that are also responsible for pulse to pulse variability and nulling in pulsars. If we assume our model applies to PSR~B0611+22 as well, one can attribute the fast mode changes to periodic shrinkage/expansion of the neutron magnetosphere. One difference would be that the magnitude of shrinkage/expansion of the magnetosphere will be small compared to what is expected in PSR~B0919+06 and PSR~B1859+07 as the offset in the emission phase between the two modes in PSR~B0611+22 is less by an order magnitude compared to PSR~B0919+06 and PSR~B1859+07. The small magnitude of phase offset could account for the extremely small timescale for the transition of the mode compared to what is observed in PSR~B0919+07 and PSR~B1859+07.

\section{Conclusions}
\label{sum}
In summary, we have performed simultaneous, broadband radio observations of PSR~B0919+06 and PSR~B1859+07 using the uGMRT and the Lovell Telescope. We sampled multiple swooshing events from both these pulsars, and this large dataset gave us an opportunity to study the properties of swooshes over a wide frequency range. We confirm previous claims that the magnitude of the swoosh is frequency dependent and pulses during a swoosh get weaker at lower frequency while pulses during the swoosh are much brighter than the normal emission at higher frequencies. We also show that the duration of swooshes is independent of the observing frequency.  Our observations favour a magnetospheric origin of the observed swooshes rather than propagation effects that can also cause chromatic emission characteristics that are commonly seen in pulsars. We invoke an emission model whereby the magnetosphere of the neutron star shrinks or expands resulting in a different line of sight traverse that can be perceived as a swoosh. We applied the model to our data to show that it can reproduce the folded profiles that are observed in our data. We also propose that the same emission model can be used to explain the radio emission in other pulsars like PSR~B0611+22 where mode changes on short timescales are observed. More multi-frequency radio observations with polarisation information of a sample of these pulsars will help us gain more insight into these phenomena.

\section*{Acknowledgements}
The authors would like to thank the anonymous reviewer whose comments greatly improved the manuscript. KMR and BWS acknowledge support from the European Research Council (ERC) under the European Union's Horizon 2020 research and innovation programme (grant agreement No 694745). JMR acknowledges support from the U.S National Science Foundation grants 09-68296 and 18-14397. KMR would like to thank Vincent Morello for useful discussion regarding the estimation of uncertainties of the duration of swooshes. KMR would like to thank Patrick Weltevrede for useful discussions and comments on the emission models used on the paper. Pulsar research at Jodrell Bank Centre for Astrophysics and Jodrell Bank Observatory is supported by a consolidated grant from the UK Science and Technology Facilities Council (STFC). We acknowledge the Department of Atomic Energy, Government of India, under project no. 12-\&D-TFR-5.02-0700 for supporting GMRT operations. The GMRT is run by the National Centre for Radio Astrophysics of the Tata Institute of Fundamental Research, India. We acknowledge support of GMRT telescope operators for observations. The Arecibo Observatory is operated by the University of Central Florida, Ana G. Mendez-Universidad Metropolitana, and Yang Enterprises under a cooperative agreement with the National Science Foundation (NSF; AST-1744119). The authors would like to thank the staff of the Lovell telescope, GMRT and Arecibo telescope for help in carrying out the observations.

\section*{Data availability}
The data will be made available to others upon reasonable request to the authors.



\bibliographystyle{mnras}
\bibliography{main} 




\bsp	
\label{lastpage}
\end{document}